\documentclass[Physsubmission, Phys]{SciPost}

\hypersetup{
    colorlinks,
    linkcolor={red!50!black},
    citecolor={blue!50!black},
    urlcolor={blue!80!black}
}

\usepackage[bitstream-charter]{mathdesign}
\DeclareSymbolFont{usualmathcal}{OMS}{cmsy}{m}{n}
\DeclareSymbolFontAlphabet{\mathcal}{usualmathcal}

\begin{document}

\begin{center}{\Large \textbf{
Longitudinal Z-boson polarization and the Higgs boson production cross-section\\
}}\end{center}

\begin{center}
  S. Amoroso\textsuperscript{1$\star$}, for the {\tt{xFitter}} Collaboration
\end{center}

\begin{center}
{\bf 1} Deutsches Elektronen-Synchrotron DESY, D 22607, Hamburg

* simone.amoroso@desy.de
\end{center}

\begin{center}
\today
\end{center}


\definecolor{palegray}{gray}{0.95}
\begin{center}
\colorbox{palegray}{
  \begin{tabular}{rr}
  \begin{minipage}{0.1\textwidth}
    \includegraphics[width=22mm]{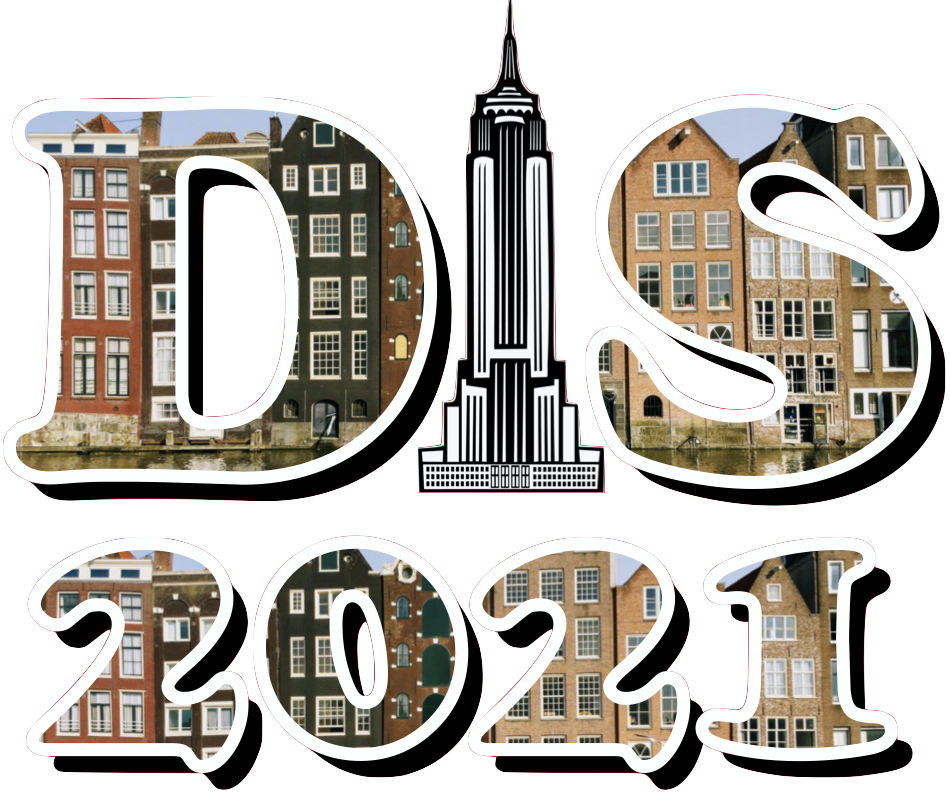}
  \end{minipage}
  &
  \begin{minipage}{0.75\textwidth}
    \begin{center}
    {\it Proceedings for the XXVIII International Workshop\\ on Deep-Inelastic Scattering and
Related Subjects,}\\
    {\it Stony Brook University, New York, USA, 12-16 April 2021} \\
    \doi{10.21468/SciPostPhysProc.?}\\
    \end{center}
  \end{minipage}
\end{tabular}
}
\end{center}

\section*{Abstract}
{\bf
  We present a study of the  LHC (and HL-LHC) potential towards a precise determination of the gluon parton distribution function
  of the proton at intermediate Bjorken-$x$ from measurements of  Drell-Yan production.
  To this extent, we exploit a clean and theoretically well predicted observable:
  the Drell-Yan lepton angular coefficient $A_0$, associated with the longitudinal  polarization of the Z boson.
  Through a detailed numerical analysis using the open-source {\tt{xFitter}} platform,
  we illustrate how this observable can provide significant sensitivity over current determinations of the gluon PDF,
  and a reduction in the PDF uncertainty on the Higgs boson production cross-section by a factor of over 50\%.
}

\section{Introduction}
\label{sec:intro}

The characterisation of the Higgs sector and the investigation of the electroweak symmetry breaking mechanism
are among the main physics goals of the LHC and future High Luminosity (HL) LHC.
The uncertainties from the knowledge of the parton distribution functions of the proton constitute
 one of the main limitations to fully exploit its physics potential~\cite{Cepeda:2019klc}.
In particular, the Higgs cross-section in its dominant gluon-gluon fusion production channel
is extremely sensitive to the gluon PDF at intermediate Bjorken-$x$ ranges.
Currently, the gluon PDF is determined indirectly from scaling violations in deep-inelastic scattering data,
with additional sensitivity coming from LHC measurements such as jet and top quark production.
These latter two processes are however challenging to measure,  and suffer from theoretical ambiguities
making them difficult to predict to high accuracy~\cite{AbdulKhalek:2020jut,Bailey:2019yze}.
In this work~\cite{Amoroso:2020fjw}, we propose a novel determination of the gluon PDF
at intermediate Bjorken-$x$ values using neutral current Drell-Yan data
and an experimentally clean and easy to measure observable,
which can be predicted theoretically to very high accuracy and exhibits a good perturbative stability.
The analysis has been performed using {\tt{xFitter}}~\cite{Alekhin:2014irh,xFitterDevelopersTeam:2017xal},
an open-source package allowing to extract PDFs and other QCD parameters.
It implements a large number of different datasets, and has been used in a variety of QCD studies;
most recently  an extraction of the pion PDF~\cite{Novikov:2020snp}
and a study of the impact of forward-backward asymmetries in Drell-Yan on PDFs~\cite{Accomando:2019vqt}.

\section{The $A_0$ angular coefficient}
We consider Drell-Yan production~\cite{Drell:1970wh} via $Z/\gamma^*$ boson exchange at the LHC.
The  cross section summed over the electroweak boson polarizations
has the angular distribution $1 + \cos^2 \theta$ and is sensitive to the gluon PDF for finite $p_T$.
However, in the small-$p_T$ region where the cross-section is the largest it is sensitive to large logarithmic corrections,
while at high-$p_T$ the missing higher order uncertainties become large.
In this study we focus on the leptonic angular distributions through the angular coefficient $A_0$,
the ratio of the longitudinal electroweak boson cross section to the unpolarized cross section:
\begin{equation}
\label{A0def}
A_0 ( s , M, Y, p_T) = { {2 d \sigma^{(L)} / dM dY dp_T } \over { d \sigma / dM dY dp_T} } .
\end{equation}
It associated with the $(3 \cos^2 \theta -1) /2$ angular dependence of the cross-section on
the polar angle $\theta^*$ ~\cite{Mirkes:1994dp}, with $\theta^*$ measured in the Collins-Soper frame~\cite{Collins:1977iv}.
The longitudinally polarized coefficient $A_0$ in Eq.~(\ref{A0def}) vanishes in the parton model and
receives leading-order (LO) perturbative QCD contributions at ${\cal O} (\alpha_s)$.
It has been computed up to ${\cal O} (\alpha_S^3)$ ~\cite{Gauld:2017tww},
showing a very good  pertubative stability.
The $A_0$ coefficient is parity-conserving and sensitive to flavor singlet PDFs.
The main sensitivity to the gluon distribution arises
from the region where the change of $A_0$ with $p_T$ is largest,
i.e., around the turn-over point $\partial^2 A_0 / \partial p_T^2=0$.
For dilepton masses near the $Z$-boson peak the turn-over
occurs for boson $p_T$ of the order of several ten to 100~GeVs.

The angular coefficients in Drell-Yan, and $A_0$, have been measured by many fixed-target and collider experiments.
We consider the $A_0$ distribution from the ATLAS measurement of the angular coefficients in Z boson production at $\sqrt{s}=8$~TeV~~\cite{ATLAS:2016rnf}.
The measurement is performed in  23 $p_T$ bins (we remove the first bins where our fixed-order predictions would not be adequate),
going from 11.4~GeV to 600~GeV, and in three different rapidity bins: $0<|Y| <1$, $1<|Y|<2$, and  $2<|Y| <3.5$.
Predictions at NLO in QCD for finite boson $p_T$ (at order ${\cal O} (\alpha^2)$) have been produced with mg5\_aMC@NLO and aMCfast~\cite{Alwall:2014hca}.
A $\chi^2$ is constructed, incorporating all experimental uncertainties and their correlations, as well as PDF uncertainties.
The impact of the $A_0$ data on PDFs is then evaluated through
an Hessian profiling technique~\cite{Paukkunen:2014zia,HERAFitterdevelopersTeam:2015cre}. 
We find a good description of the data, with $\chi^2$/ndf of order one,
for all modern PDFs: CT18NNLO~\cite{Hou:2019efy}, NNPDF3.1nnlo~\cite{NNPDF:2017mvq}, ABMP16nnlo~\cite{Alekhin:2017kpj}, and MSHT20nnlo~\cite{Bailey:2020ooq}.
The reduction in the PDF uncertainties after the inclusion of the $A_0$ data is however found to be negligible.






\section{PDF constraints and the Higgs cross-section}
\label{sec:another}
\subsection{Sensitivity to the gluon PDF}
\begin{figure}[h]
  \centering
  \includegraphics[width=0.45\textwidth]{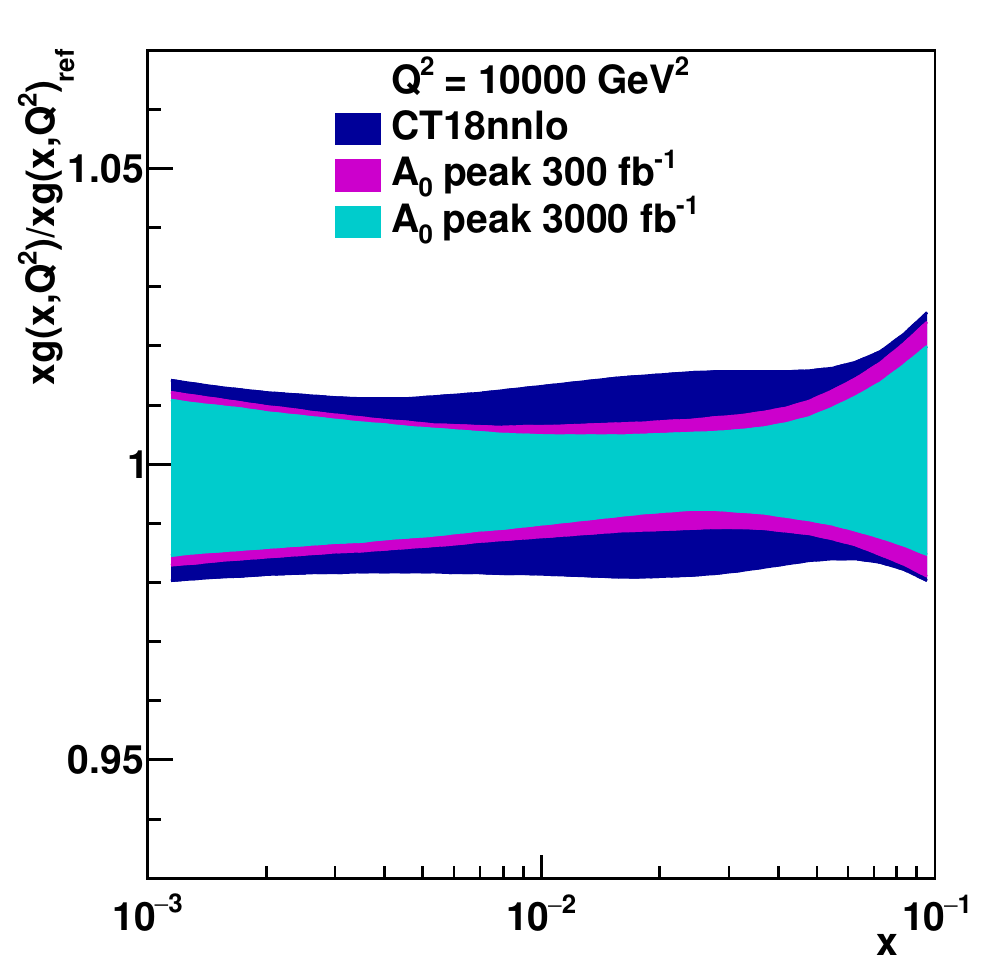}
  \includegraphics[width=0.45\textwidth]{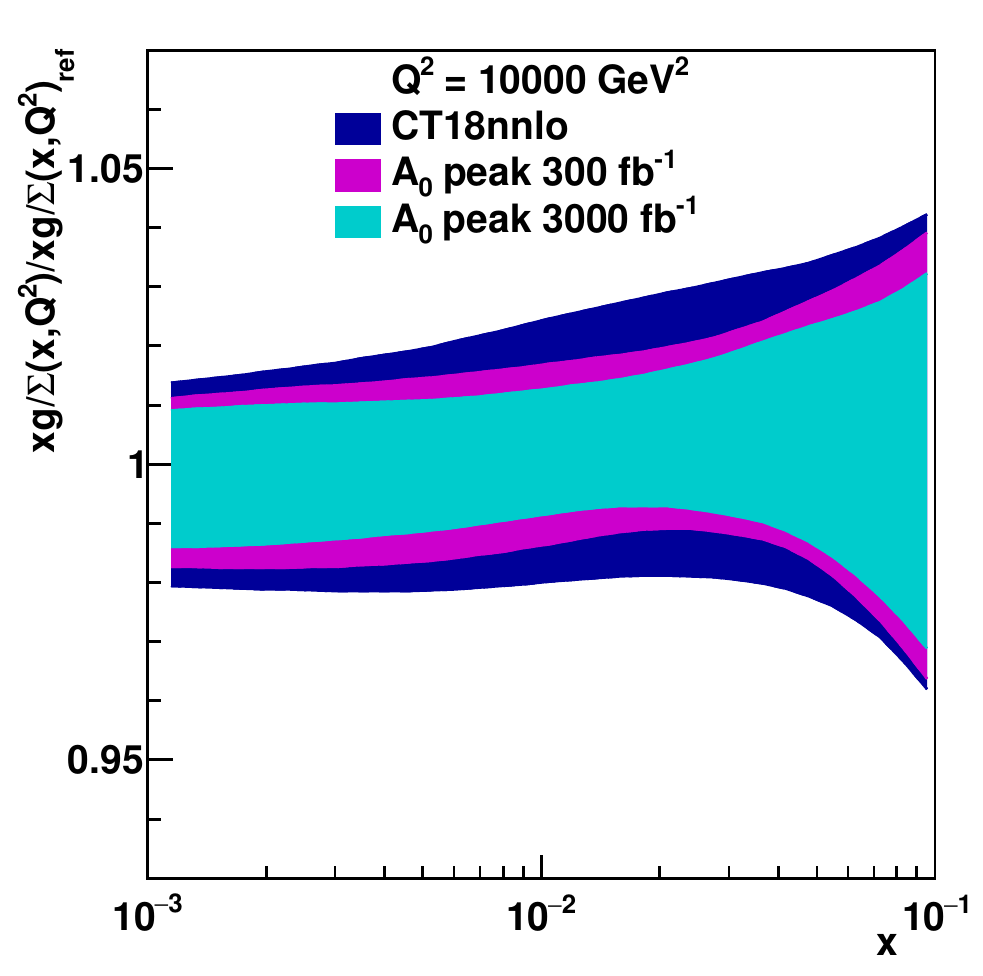}
  \caption{The gluon ($xg$) and gluon/Sea ($xg/\Sigma$) PDFs in the  original CT18NNLO analysis (red) and profiled using $A_0$ pseudodata corresponding
    to integrated luminosities of 300~fb$^{-1}$ (blue) and 3~ab$^{-1}$ (green) in the region 80~GeV $< M <$ 100~GeV and $|Y|<3.5$.}
  \label{fig:pdf}
\end{figure}

We then generate  pseudodata for the $A_0$ distribution at $\sqrt{s} = 13$~TeV for two projected
luminosity scenarios of 300 fb$^{-1}$ (the expected integrated luminosity at the end of the LHC Run III)
and 3 ab$^{-1}$ (the design integrated luminosity  of the HL-LHC stage). 
To do this we extrapolate the statistical uncertainties, and add a conservative estimate
of 0.1\% for the dominant experimental systematic  on the lepton momentum scale.
We perform a PDF profiling exercise using the CT18NNLO PDF set, first reduced to 68\% CL coverage,
and find that the  increased statistics allows for a significant reduction in the PDF uncertainties.
This reduction is  particularly large for  the gluon density and for the $u$ and $d$
sea-quark densities coupled to gluons through QCD evolution, as shown in Fig.~\ref{fig:pdf}.
The largest sensitivity is found to come from transverse momenta in the range $p_T \sim 50$~GeV, dying out for $p_T > 100$~GeV.
The gain from a 300 fb$^{-1}$  measurement is found to dominate over the 3~ab$^{-1}$ gain,
although an improvement is nonetheless present.

\subsection{Impact on Higgs phenomenology}
\begin{figure}[h]
\centering
\includegraphics[width=0.48\textwidth]{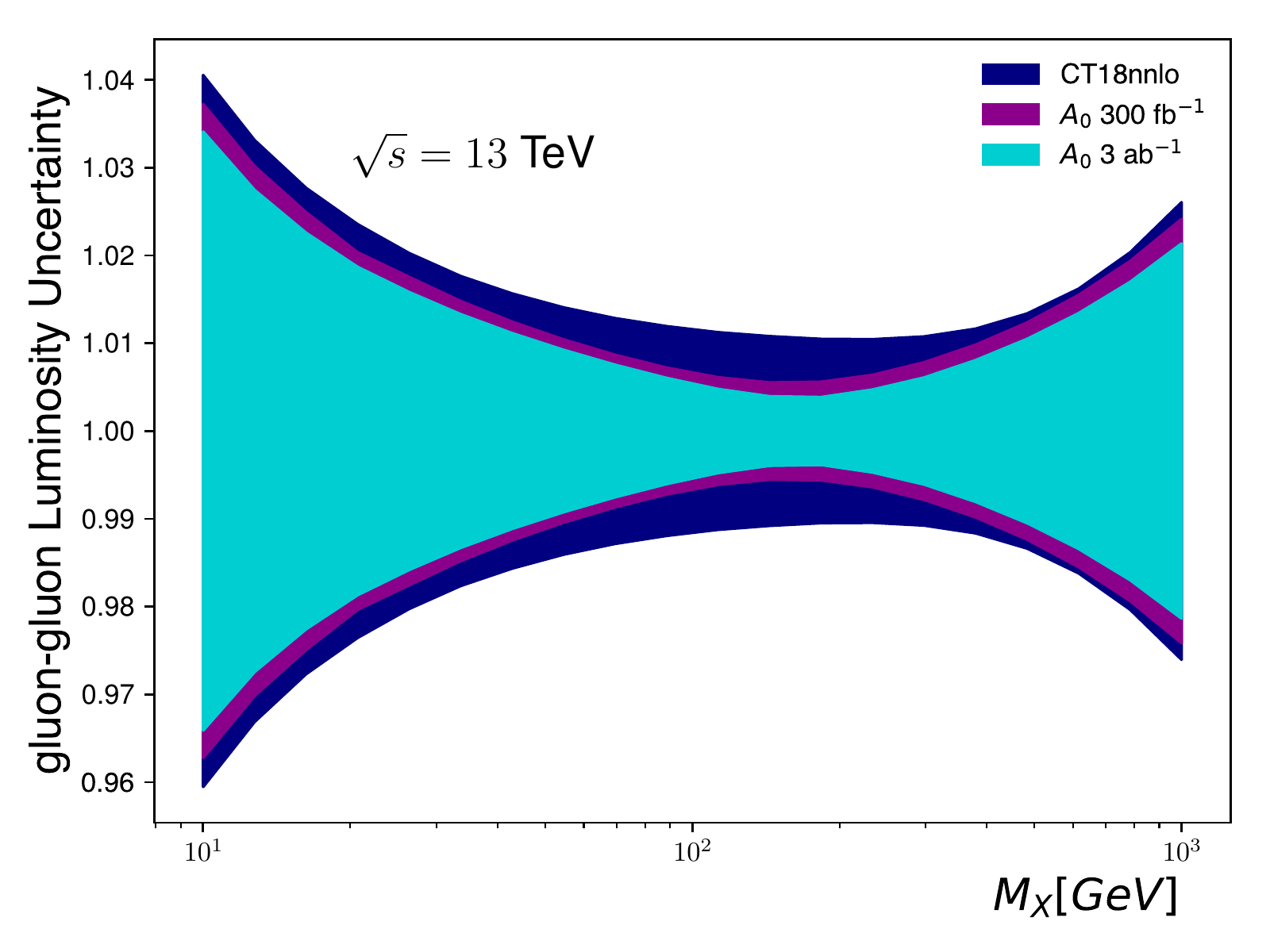}
\includegraphics[width=0.48\textwidth]{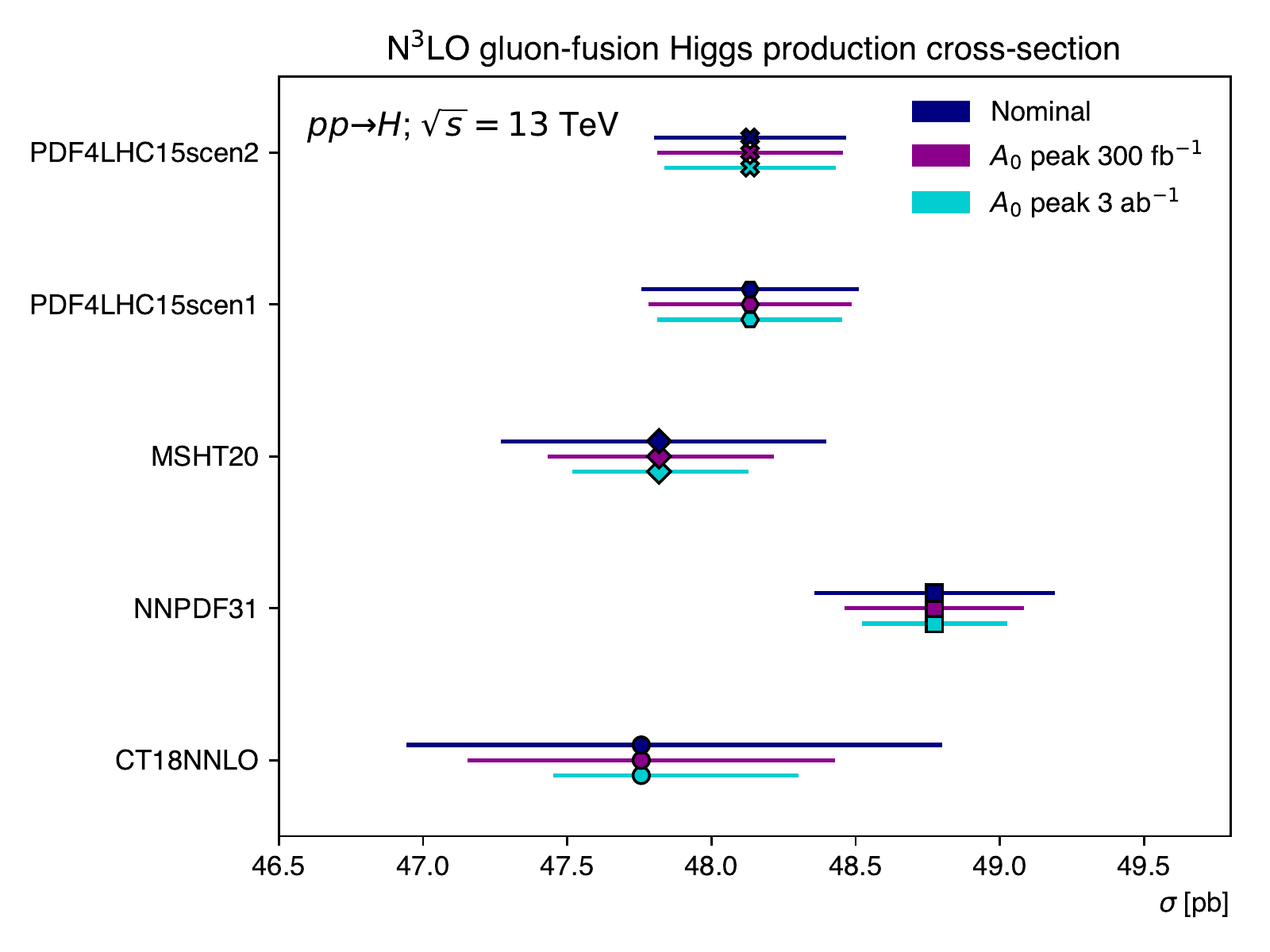}
\caption{[Left] Ratio of PDF uncertainties for the gluon-gluon luminosity evaluated at $\sqrt{s}=$13~TeV.
  The red band shows the uncertainties of the CT18NNLO PDF set~\cite{Hou:2019efy}, reduced to 68\% CL coverage.
  The blue and green bands show the impact of including constraints
  from the $A_0$ measurement and assuming 300~fb$^{-1}$ and 3~ab$^{-1}$, respectively.
  [Right] PDF uncertainties on the N3LO  gluon-gluon fusion Higgs boson cross-section, 
  shown for different PDF sets before and after the inclusion of the $A_0$ measurement constraints.}
\label{fig:higgs}
\end{figure}

The effect of the longitudinally polarized coefficient on the $Q^2$ = 10$^4$ GeV$^2$ gluon PDF near $ x \sim 10^{-2}$
will influence the Higgs boson cross section.
In Fig.~\ref{fig:higgs} we show the gluon-gluon luminosity for the CT18NNLO PDF,
before and after the profiling using $A_0$ pseudodata.
Most of the  reduction in the uncertainties is found  in the Higgs mass range.
In the right plot we show the N$^3$LO  gluon-fusion Higgs total cross section and its uncertainty
obtained with  {\tt{ggHiggs}}~\cite{Bonvini:2014jma,Bonvini:2018ixe}.
for the CT18NNLO, NNPDF3.1nnlo and MSHT20nnlo global PDF sets.
Notwithstanding the numerical differences,
a similar reduction in the PDF uncertainties is seen for the different sets.
We also show The PDF4LHC15scen1 and 2 projected sets,
which  include  pseudodata based on complete LHC data sample~\cite{Khalek:2018rok},
and which show a smaller, but not negligible, reduction in uncertainties.

\section{Conclusion}
We have studied the impact of precise measurements of the $A_0$ angular coefficient in neutral current Drell-Yan
on the uncertainties of the gluon PDF and on the Higgs boson production cross-section in gluon-gluon fusion.
We found  that future  measurements of $A_0$ have the possibility to reduce by about a factor of two
the uncertainty of the gluon PDFs at intermediate Bjorken-$x$ values.
This translates in a large reduction in the PDF uncertainty
on the gluon-fusion Higgs production cross-section.
Our results open a new area of phenomenological studies
on connections between the gauge and Higgs sectors of the SM.
Further aspects may be investigated by extending the analysis
to the full structure of lepton angular distributions, including polarization interferences,
and to mass regions far away from the $Z$-boson peak.

 \bibliographystyle{SciPost_bibstyle} 
\bibliography{SciPost_Latex_Template.bib}

\begin{thebibliography}{10}
\providecommand{\url}[1]{\texttt{#1}}
\providecommand{\urlprefix}{URL }
\expandafter\ifx\csname urlstyle\endcsname\relax
  \providecommand{\doi}[1]{doi:\discretionary{}{}{}#1}\else
  \providecommand{\doi}{doi:\discretionary{}{}{}\begingroup
  \urlstyle{rm}\Url}\fi
\providecommand{\eprint}[2][]{\url{#2}}

\bibitem{Cepeda:2019klc}
M.~Cepeda \emph{et~al.},
\newblock \emph{{Report from Working Group 2}: {Higgs Physics at the HL-LHC and
  HE-LHC}},
\newblock CERN Yellow Rep. Monogr. \textbf{7}, 221 (2019),
\newblock \doi{10.23731/CYRM-2019-007.221},
\newblock \eprint{1902.00134}.

\bibitem{AbdulKhalek:2020jut}
R.~Abdul~Khalek \emph{et~al.},
\newblock \emph{{Phenomenology of NNLO jet production at the LHC and its impact
  on parton distributions}},
\newblock Eur. Phys. J. C \textbf{80}(8), 797 (2020),
\newblock \doi{10.1140/epjc/s10052-020-8328-5},
\newblock \eprint{2005.11327}.

\bibitem{Bailey:2019yze}
S.~Bailey and L.~Harland-Lang,
\newblock \emph{{Differential Top Quark Pair Production at the LHC: Challenges
  for PDF Fits}},
\newblock Eur. Phys. J. C \textbf{80}(1), 60 (2020),
\newblock \doi{10.1140/epjc/s10052-020-7633-3},
\newblock \eprint{1909.10541}.

\bibitem{Amoroso:2020fjw}
S.~Amoroso, J.~Fiaschi, F.~Giuli, A.~Glazov, F.~Hautmann and O.~Zenaiev,
\newblock \emph{{Longitudinal Z-Boson Polarization and the Higgs Boson
  Production Cross Section at the Large Hadron Collider}}  (2020),
\newblock \eprint{2012.10298}.

\bibitem{Alekhin:2014irh}
S.~Alekhin \emph{et~al.},
\newblock \emph{{HERAFitter}},
\newblock Eur. Phys. J. C \textbf{75}(7), 304 (2015),
\newblock \doi{10.1140/epjc/s10052-015-3480-z},
\newblock \eprint{1410.4412}.

\bibitem{xFitterDevelopersTeam:2017xal}
V.~Bertone \emph{et~al.},
\newblock \emph{{xFitter 2.0.0: An Open Source QCD Fit Framework}},
\newblock PoS \textbf{DIS2017}, 203 (2018),
\newblock \doi{10.22323/1.297.0203},
\newblock \eprint{1709.01151}.

\bibitem{Novikov:2020snp}
I.~Novikov \emph{et~al.},
\newblock \emph{{Parton Distribution Functions of the Charged Pion Within The
  xFitter Framework}},
\newblock Phys. Rev. D \textbf{102}(1), 014040 (2020),
\newblock \doi{10.1103/PhysRevD.102.014040},
\newblock \eprint{2002.02902}.

\bibitem{Accomando:2019vqt}
E.~Accomando \emph{et~al.},
\newblock \emph{{PDF Profiling Using the Forward-Backward Asymmetry in Neutral
  Current Drell-Yan Production}},
\newblock JHEP \textbf{10}, 176 (2019),
\newblock \doi{10.1007/JHEP10(2019)176},
\newblock \eprint{1907.07727}.

\bibitem{Drell:1970wh}
S.~D. Drell and T.-M. Yan,
\newblock \emph{{Massive Lepton Pair Production in Hadron-Hadron Collisions at
  High-Energies}},
\newblock Phys. Rev. Lett. \textbf{25}, 316 (1970),
\newblock \doi{10.1103/PhysRevLett.25.316},
\newblock [Erratum: Phys.Rev.Lett. 25, 902 (1970)].

\bibitem{Mirkes:1994dp}
E.~Mirkes and J.~Ohnemus,
\newblock \emph{{Angular distributions of Drell-Yan lepton pairs at the
  Tevatron: Order $\alpha-s^{2}$ corrections and Monte Carlo studies}},
\newblock Phys. Rev. D \textbf{51}, 4891 (1995),
\newblock \doi{10.1103/PhysRevD.51.4891},
\newblock \eprint{hep-ph/9412289}.

\bibitem{Collins:1977iv}
J.~C. Collins and D.~E. Soper,
\newblock \emph{{Angular Distribution of Dileptons in High-Energy Hadron
  Collisions}},
\newblock Phys. Rev. D \textbf{16}, 2219 (1977),
\newblock \doi{10.1103/PhysRevD.16.2219}.

\bibitem{Gauld:2017tww}
R.~Gauld, A.~Gehrmann-De~Ridder, T.~Gehrmann, E.~W.~N. Glover and A.~Huss,
\newblock \emph{{Precise predictions for the angular coefficients in Z-boson
  production at the LHC}},
\newblock JHEP \textbf{11}, 003 (2017),
\newblock \doi{10.1007/JHEP11(2017)003},
\newblock \eprint{1708.00008}.

\bibitem{ATLAS:2016rnf}
G.~Aad \emph{et~al.},
\newblock \emph{{Measurement of the angular coefficients in $Z$-boson events
  using electron and muon pairs from data taken at $\sqrt{s}=8$ TeV with the
  ATLAS detector}},
\newblock JHEP \textbf{08}, 159 (2016),
\newblock \doi{10.1007/JHEP08(2016)159},
\newblock \eprint{1606.00689}.

\bibitem{Alwall:2014hca}
J.~Alwall, R.~Frederix, S.~Frixione, V.~Hirschi, F.~Maltoni, O.~Mattelaer,
  H.~S. Shao, T.~Stelzer, P.~Torrielli and M.~Zaro,
\newblock \emph{{The automated computation of tree-level and next-to-leading
  order differential cross sections, and their matching to parton shower
  simulations}},
\newblock JHEP \textbf{07}, 079 (2014),
\newblock \doi{10.1007/JHEP07(2014)079},
\newblock \eprint{1405.0301}.

\bibitem{Paukkunen:2014zia}
H.~Paukkunen and P.~Zurita,
\newblock \emph{{PDF reweighting in the Hessian matrix approach}},
\newblock JHEP \textbf{12}, 100 (2014),
\newblock \doi{10.1007/JHEP12(2014)100},
\newblock \eprint{1402.6623}.

\bibitem{HERAFitterdevelopersTeam:2015cre}
S.~Camarda \emph{et~al.},
\newblock \emph{{QCD analysis of $W$- and $Z$-boson production at Tevatron}},
\newblock Eur. Phys. J. C \textbf{75}(9), 458 (2015),
\newblock \doi{10.1140/epjc/s10052-015-3655-7},
\newblock \eprint{1503.05221}.

\bibitem{Hou:2019efy}
T.-J. Hou \emph{et~al.},
\newblock \emph{{New CTEQ global analysis of quantum chromodynamics with
  high-precision data from the LHC}},
\newblock Phys. Rev. D \textbf{103}(1), 014013 (2021),
\newblock \doi{10.1103/PhysRevD.103.014013},
\newblock \eprint{1912.10053}.

\bibitem{NNPDF:2017mvq}
R.~D. Ball \emph{et~al.},
\newblock \emph{{Parton distributions from high-precision collider data}},
\newblock Eur. Phys. J. C \textbf{77}(10), 663 (2017),
\newblock \doi{10.1140/epjc/s10052-017-5199-5},
\newblock \eprint{1706.00428}.

\bibitem{Alekhin:2017kpj}
S.~Alekhin, J.~Bl\"umlein, S.~Moch and R.~Placakyte,
\newblock \emph{{Parton distribution functions, $\alpha_s$, and heavy-quark
  masses for LHC Run II}},
\newblock Phys. Rev. D \textbf{96}(1), 014011 (2017),
\newblock \doi{10.1103/PhysRevD.96.014011},
\newblock \eprint{1701.05838}.

\bibitem{Bailey:2020ooq}
S.~Bailey, T.~Cridge, L.~A. Harland-Lang, A.~D. Martin and R.~S. Thorne,
\newblock \emph{{Parton distributions from LHC, HERA, Tevatron and fixed target
  data: MSHT20 PDFs}},
\newblock Eur. Phys. J. C \textbf{81}(4), 341 (2021),
\newblock \doi{10.1140/epjc/s10052-021-09057-0},
\newblock \eprint{2012.04684}.

\bibitem{Bonvini:2014jma}
M.~Bonvini, R.~D. Ball, S.~Forte, S.~Marzani and G.~Ridolfi,
\newblock \emph{{Updated Higgs cross section at approximate N$^3$LO}},
\newblock J. Phys. G \textbf{41}, 095002 (2014),
\newblock \doi{10.1088/0954-3899/41/9/095002},
\newblock \eprint{1404.3204}.

\bibitem{Bonvini:2018ixe}
M.~Bonvini and S.~Marzani,
\newblock \emph{{Double resummation for Higgs production}},
\newblock Phys. Rev. Lett. \textbf{120}(20), 202003 (2018),
\newblock \doi{10.1103/PhysRevLett.120.202003},
\newblock \eprint{1802.07758}.

\bibitem{Khalek:2018rok}
R.~Abdul~Khalek, S.~Bailey, J.~Gao, L.~Harland-Lang and J.~Rojo,
\newblock \emph{{Towards Ultimate Parton Distributions at the High-Luminosity
  LHC}},
\newblock Eur. Phys. J. C \textbf{78}(11), 962 (2018),
\newblock \doi{10.1140/epjc/s10052-018-6448-y},
\newblock \eprint{1810.03639}.

\end{thebibliography}

\nolinenumbers

\end{document}